\documentstyle[aps]{revtex}
\title{NARROWING OF BENNETT HOLE\\
IN COLLISIONAL  PLASMA}
\author{E.V.Podivilov, D.A.Shapiro, M.G.Stepanov}
\address{Institute of Automation \& Electrometry,\\
Novosibirsk, 630090, Russia}

\begin{document}
\draft
\maketitle
\begin{abstract}
The profile of a Bennett hole induced by laser field in ionic
distribution in collisional plasma is calculated. Influence of
Chandrasekhar's dependence of coefficients of velocity space
transport on the profile is included into the calculation for the
first time. It is found that the hole narrows down as the field
detuning frequency increases.  Physical cause of the effect is the
falling dependence of Coulomb collision frequency on the ionic
velocity.  Estimations show that the effect is quite observable under
conditions of high-current gas-discharge plasma.

\pacs{52.20Hv, 42.62Fi, 52.80.-s}

\end{abstract}

The nonlinear spectroscopic technique as a plasma diagnostic tool is
a subject of considerable recent interest because of its novelty and
potential in basic and applied plasma physics. Of particular
interest are the spectroscopic effects of Coulomb scattering
\cite{Babin}, since it differs radically in appearance from binary
collisions in gases \cite{Berman,Shalagin}. In linear absorption
spectra
Doppler width usually exceeds collisional broadening, so it masks
effects of the scattering. However, the velocity distribution of
particles interacting with a monochromatic field near the resonance
is changed significantly by the scattering. If the frequency $\omega
$ of the field is near-resonant with Bohr's frequency $\omega
_{mn}=(E_m-E_n)/\hbar $ between levels $m$ and $n$ of the atoms with
velocity $\vec v$ that satisfies the condition $\vec k\vec v=\omega
-\omega _{mn}\equiv \Omega $, then the induced transitions
$m\leftrightarrow n$ occur ($\vec k$ denotes the wavevector). As a
result, the narrow peak or dip (Bennett hole \cite{Bennett62}) arises
in the velocity distribution of level populations. The steady-state
profiles of holes are controlled by both the homogeneous width of
transition $m\leftrightarrow n$ and the scattering processes with a
change in velocity. Measuring the width, shift and asymmetry of
nonequilibrium structures in the distribution with the help of
probe-field technique one can extract information about the
scattering.

Previous calculations of Coulomb broadening \cite{Babin,Smirnov} were
based on the assumption of the constancy of the diffusion coefficient
in the velocity space. This assumption is applicable when the
detuning frequency $\Omega $ is less than the Doppler width $kv_T$,
where $v_T=\sqrt{2T_i/m}$ is the thermal velocity of ions and $m$ is
the ion mass. The constant diffusion model was enough to interpret
Lamb dip measurement in argon laser plasma or another with near-
resonance
pumping. In contrast, farther away from the resonance, at $\Omega
>kv_T$, one must take Chandrasekhar's velocity dependence of the
diffusion tensor that complicates the evaluation. One way
around this problem was shown in linear theory of
Dicke narrowing in the ionic spectrum, which had been built up for
low intensity of the incident wave \cite{Chernykh}.

In the present Letter we study a first-order nonlinear problem, the
influence of collisions in plasma on the profile of the Bennett hole.
The calculation presented below for nonlinear structure in the
velocity distribution is performed in the basis of density matrix
formalism \cite{Shalagin,Stenholm}. First we derive the formula for
the distribution, then analyze it under some simple but realistic
conditions, and at the end estimate whether the effect obtained be
observable in experiment.  In particular, we show that the width of
hole decreases with the detuning of the electromagnetic field from
the resonance. By observing the effect, one could directly
measure the diffusion tensor. It would be helpful to confirm our
understanding and offer scope for new diagnostic methods.

The diffusion coefficient as a function of the velocity was measured
for
Q-machine plasma at low density by Bowles et al. \cite{Bowles}. The
effect
of transient spread of a Bennett hole in the distribution of tagged
ion
population in a metastable state was exploited. However, that effect
cannot
be used for denser plasma, because the diffusion time becomes smaller
and
the measurement requires a faster registration technique. In this
paper we
shall examine the stationary hole shape. Analysis of its width is
useful for
diagnostics at high as well as at low density.

To develop the theory of nonequilibrium structures in the
velocity distribution, let us describe the two-level subsystem of a
probe ion in an ideal nondegenerate plasma by the spectroscopic
density matrix $\rho _{ij}( \vec r,\vec v,t)$. This
matrix over internal states $i,j$ is at the same time the Wigner
function of position $\vec r$ and velocity $\vec v$. It has been
shown \cite{Smirnov,Chernykh} that elements of the matrix satisfy the
quantum kinetic equation with the classical Landau collision term.
This
equation makes it possible to analyze resonant transitions between
levels in
the field of the traveling wave $\vec E=\vec E_0
\exp (-i\omega t+i\vec k\vec r)+$ c.c.

Denoting the relaxation constants of level populations as $\Gamma_j$
and that of the coherence as $\Gamma $, write steady - state
equations for off-diagonal $\rho_{mn}(\vec r,\vec v,t)=\rho (\vec
v)\exp (-i\Omega t+i\vec k\vec r)$ and diagonal $\rho _{jj}\equiv
\rho _j$ elements of the density matrix
\widetext
\begin{equation}\label{qke}
\left(\Gamma - i \Omega + i \vec k\vec v\right)\rho -\nu
\hat V \rho  = - iG \Delta N,\quad
\Gamma_j\rho_j -\nu \hat V  \rho_j - \lambda_j(\vec v) = \mp
2 \Re (iG^*\rho),
\end{equation}
where the upper/lower sign corresponds to upper/lower level. Here
$\Delta
N\equiv \rho _m-\rho _n$, $G=\vec E_0\vec d_{mn}/2\hbar $,
$\vec d_{mn}$ is the matrix element of the dipole moment,
\begin{equation}
\label{Nu}\nu ={\frac{16\sqrt{\pi }LN_i(Z_ae^2)^2}{3m^2v_T^3}}
\end{equation}
is the Coulomb collision frequency, $Z_ae$ is the charge of a probe
ion, $L$
is the Coulomb logarithm. We assume that there are only singly
charged ions
in plasm; $N_i$ is their concentration.
The excitation of ionic levels $j=m,n$ usually occurs from the ground
or
metastable state, so we are justified in assuming the shape of the
excitation function $\lambda _j(\vec v)$ to be Maxwellian
\begin{equation}
\lambda _j(\vec v)=Q_jW(\vec v),\quad W(\vec v)
={\frac 1{v_T^3\pi ^{3/2}}}\exp \left( -{\frac{v^2}{v_T^2}}\right) ,
\end{equation}
where $Q_j$ is the excitation rate of level $j$.

For plasma in a state close to the thermodynamic equilibrium with
ionic
temperature $T_i$, the distribution of buffer particles has the
Maxwellian shape. Then it is possible to write an explicit expression
for
collision operator $\hat V$ involving the dynamic friction and the
diffusion
in the velocity space
\begin{equation}
\label{Stos}
\hat V={\frac 12}{\frac \partial {\partial \xi _\alpha }}\Phi
_{\alpha \beta }\left( {\frac \partial {\partial \xi _\beta }}+2\xi
_\beta
\right),
\end{equation}
where $\xi _\alpha =v_\alpha /v_T$ is the $\alpha $ - component of the
dimensionless velocity, the differential operator $\partial /\partial
\xi
_\alpha $ acts on the right. Functions $\Phi _l(\xi )$ and
$\Phi _{tr}(\xi )$, occurring in the expression for the diffusion
tensor
\begin{equation}
\label{Tensor}\Phi _{\alpha \beta }=\Phi _l(\xi )\frac{\xi _\alpha \xi
_\beta }{\xi ^2}+\Phi _{tr}(\xi )\left( \delta _{\alpha \beta }-
\frac{\xi
_\alpha \xi _\beta }{\xi ^2}\right) ,
\end{equation}
can be written as single integrals
\begin{equation}
\label{Velocity dependence}\Phi _l(\xi )=3\int\limits_0^1\lambda ^2
e^{-\lambda ^2\xi ^2}\,d\lambda ,\quad \Phi _{tr}(\xi )=
{\frac 32}\int\limits_0^1(1-\lambda ^2)e^{-\lambda ^2\xi
^2}\,d\lambda.
\end{equation}

The velocity distribution of populations $\rho _j$ of both levels
$j=m,n$
needs to be found. To simplify analysis, we assume the electromagnetic
field to be weak $|G|^2\ll \Gamma \Gamma _j$ and will neglect the
effects of
strong saturation. The right sides of Eqs. (\ref{qke}) can
therefore be regarded as perturbation.
The formal solution of (\ref{qke}) in terms of operator exponents is
$$
\rho _j^{(0)}={\frac{Q_j}{\Gamma _j}}W(\vec v),\quad \rho
^{(0)}=0;\quad \Delta N(\vec v)=\rho _m^0(\vec v)-
\rho_n^{(0)}(\vec v)=N_{mn}W(\vec v);
$$
\begin{eqnarray} &&\rho^{(1)}(\vec v) =
-iG\int\limits_0^{\infty}dt \exp(-\hat A t + \nu \hat V t)
\Delta N (\vec v),\quad
\hat A = \Gamma - i \Omega + i \vec k \vec v,\\
&&\rho_j^{(1)}(\vec v) = \mp\int\limits_0^{\infty}dt
\exp(-\Gamma_j t + \nu \hat V t)
2\Re \left(iG^* \rho^{(1)}(\vec v) \right).
\end{eqnarray}

It remains to write out the explicit solution and analyze its
limiting cases. For this purpose we must calculate the commutators of
operators $\hat A$ and $\hat V$ and get, in the general case, the
infinite
series. Fortunately, the relation $[[[\hat V,\hat A],\hat A],\hat
A]=0$ allows us to break the series within the
first order in~$\nu $:
\begin{eqnarray}
&&\rho_j^{(1)}(\vec v) =\mp 2|G|^2\Delta N(\vec v) \Re
\int\limits_0^{\infty}dt \int\limits_0^{\infty}d\tau
\exp\left(-\Gamma_j \tau - (\Gamma - i\Omega + i k
v_{z})t\right)\times\nonumber\\
&&\times\exp\left[ {\nu\over 2}\left(-
\Phi_{zz}(\vec \xi)k^2v_T^2(t^2\tau+t^3/3) +2i k v_{z}
\Phi_l(\xi)(2t\tau+t^2)\right) +{\cal O} (\nu^2) \right],
\label{Bennett}
\end{eqnarray}
where the axis $z$ is chosen along the wavevector $\vec k$. Function
\begin{equation}
\label{zz}\Phi _{zz}(\vec\xi)=\left(\Phi _l(\xi )-\Phi _{tr}(\xi
)\right){\ \frac{\xi _z^2}{\xi ^2}}+\Phi _{tr}(\xi )
\end{equation}
is the $zz$ - component of the diffusion tensor. This function
depends on two variables $\xi \equiv |\vec\xi|$ and $\xi _z$. The
integrand in (\ref{Bennett}), obtained by the expansion, is
inappropriate at long times $t,\ \tau $. Consequently, for
application of this expansion it is necessary to have the main
contribution to integral (\ref{Bennett}) in the domain where the
quadratic terms in $\nu $ remain small. Since $|\partial \Phi
_{\alpha \beta }/\ \partial \xi _\gamma |<|\Phi _{\alpha \beta }|$,
estimation of those terms gives the following applicability
conditions of the expansion (\ref{Bennett})
\begin{equation}
\label{condition}\alpha ={\frac \nu {\Gamma _j}}\ll 1,\quad \beta
={\frac
\Gamma {kv_T}}\ll 1.
\end{equation}
The latter inequality means the Doppler limit; the former one
coincides with
the condition of smallness of the diffusion broadening compared to the
Doppler width. The collision frequency in high-current gas-discharge
plasma
is about $\nu \sim 10^5$ -- $10^7$~s$^{-1}$ (see \cite{Babin}). It is
by a
few orders of magnitude less than the Doppler width $kv_T\sim
10^{10}s^{-1}$
and, as a rule, less than the relaxation rates
$\Gamma \sim 10^8-10^9s^{-1}$, $\Gamma _j\sim 10^6-10^9s^{-1}$.
That is why we may use small parameters $\alpha $, $\beta $
and it is enough to hold only terms linear in $\nu $.

Taking an integral over $\tau $ we obtain the correction to the
distribution
function as a single integral over $t$
\begin{equation}
\label{nb}\rho _j^{(1)}(\vec v)=\mp 2|G|^2\Delta N(\vec v)\Re
\int\limits_0^\infty {\exp \left[ -\hat At-\Phi _{zz}(
\vec\xi)\nu k^2v_T^2t^3/6+i\nu kv_z\Phi _l(\xi )t^2\right] \over
\Gamma _j+\Phi _{zz}(\vec\xi)\nu k^2v_T^2t^2/2-2i\nu kv_z\Phi
_l(\xi )t}\,dt.
\end{equation}
The integrand in (\ref{nb}) regarded as a function of $t$ at $\Omega
=kv_z$
is maximal at $t=0$ and decreases with $t$. The shape of the
distribution
depends on a few parameters. Considering conditions (\ref{condition})
and
the inequality $\Gamma _j\leq \Gamma $ we can set apart two limiting
cases
of parameters:
\begin{equation}
\label{D1}\nu k^2v_T^2\ll \Gamma ^2\Gamma _j\quad (\alpha \ll \beta
^2),
\end{equation}
\begin{equation}
\label{D2}\Gamma ^2\Gamma _j\ll \nu k^2v_T^2\quad (\alpha \gg \beta
^2).
\end{equation}
For the domain (\ref{D1}) integral (\ref{nb}) is accumulated at $t\leq
T_1=1/\Gamma $. The diffusion and friction cause the Bennett hole to
somewhat change its width and position. For the limit (\ref{D2})
integral
(\ref{nb}) is gathered at $t\leq T_2=(\alpha /2)^{-1/2}/kv_T$.
Diffusion and
friction effects control the width and shift of the hole. The width is
defined by the maximum among the homogeneous width $\Delta v_H=\Gamma
/k$ and diffusion width $\Delta v_D=v_T(\nu /\Gamma _j)^{1/2}$. Thus,
the scattering has stronger impact on shape of the hole, while the
relaxation constants of selected levels satisfy inequality
(\ref{D2}).

Let us find the width as a function of detuning in this limit
$\epsilon(\vec\xi)=2\beta ^2/\alpha \Phi _{zz}(\vec\xi)\ll 1$. It is
convenient to write the expression for the nonlinear correction to
the Maxwellian distribution as a convolution of two functions
\begin{equation}
\label{cb}\rho _j^{(1)}(\vec v)=W(\vec v)\int\limits_{-\infty}^\infty
f(x-x^\prime)g_j(x^{\prime })\,dx^\prime,\quad
x=kv_z-\Omega ,
\end{equation}
where
\begin{eqnarray}
g_j(x) = \mp {\pi |G|^2 N_{mn}\over \Gamma\Gamma_j}
{\epsilon\over\sqrt\mu}
\exp\left(-{|x| \sqrt{\mu}+x\sigma\over \Gamma}
\right),\label{g}\\
\sigma(\vec \xi) = {2\Gamma \xi_z \Phi_l(\xi)\over
kv_T \Phi_{zz}(\vec \xi)}, \quad \mu(\vec \xi) =
\epsilon(\vec \xi) + \sigma^2 \simeq \epsilon(\vec \xi),
\nonumber\\
f(x) = \Re\int\limits_0^{\infty}{dt\over \pi}\,
\exp\left[- (\Gamma + i x)t -
\Phi_{zz}(\vec\xi)\nu k^2v_T^2 t^3/6
+ i\nu k v_{z} \Phi_l(\xi)t^2 \right].\label{f}
\end{eqnarray}
In the limit $\epsilon \ll 1$, the width of function $f(x)$ is much
less than that of $g_j(x)$. Accordingly, the central part of the
distribution is given by expression (\ref{g}) $\rho _j^{(1)}\simeq
g_j(x)$ at $|x|<\Gamma / \sqrt{\epsilon }$. Meanwhile function
$g_j(x)$ drops exponentially with $|x|$, so the asymptotic of the
distribution $\rho _j^{(1)}(v_z)$ at $|x|\gg \Gamma /\sqrt{\epsilon
}$ is given by a narrow but slowly falling function $f(x)$ and has
the Lorentzian shape.

The most interesting feature of the distribution obtained is the
narrowing
of its central part as the detuning $\Omega $ increases, due to the
falling
velocity dependence of the collision frequency. One can estimate the
half-width by substituting the extreme value of the longitudinal
velocity
$v_z=\Omega /k$ into all components of the diffusion tensor:
$$
\Delta x\simeq {\frac \Gamma {\sqrt{\mu }}}\simeq kv_T\sqrt{\frac{\nu
\Phi
_{zz}(\xi ,\Omega /kv_T)}{\Gamma _j}}.
$$
This quantity is proportional to the square root of the $zz$-
component of
the diffusion tensor. It falls off with detuning $\Omega $. The
effect is absent in the model of the constant collision frequency
\cite{Rautian,Gelmedova}.

To describe this effect quantitatively we calculate the $\Omega $-
dependence
of $x^2$ averaged over the distribution (\ref{cb}). Keeping only the
terms
linear in parameters $\alpha $ and $\beta $ we obtain
\begin{equation}
\label{width}F_j(\Omega )=\int d^3v\,x^2
\rho_j^{(1)}(\vec v)\simeq \mp {\frac{2|G|^2N_{mn}}
{\Gamma _j}}\left\{ \Gamma +{\frac{\nu k^2v_T^2}{\Gamma _j}}
\Re\int\limits_0^{\infty} dt\,
\overline{\Phi}_{zz}(t)
\exp\left(- (\Gamma - i \Omega)t \right)
\right\},
\end{equation}
$$
\overline{\Phi }_{zz}(t)=\int d^3vW(\vec v)\Phi _{zz}(\vec\xi)
\exp (-i\xi _zt)=\int\limits_0^{1/\sqrt{2}}6y^2dy\exp
\left[-(1-y^2)\left({tkv_T\over 2}\right)^2\right].
$$
The first term on the right-hand side of (\ref{width}) is independent
of
$\Omega $ and corresponds to the contribution of the Lorentzian wings
of the
distribution. The second term describes the collisional broadening and
decreases with $|\Omega |$.

Under the approximations the line shape is Voigt's contour. The
dependence
of the variance $D=\langle (x-\langle x\rangle )^2\rangle $ on
detuning
$\Omega $ consists of two terms: the contribution of the wings, which
increases with detuning, and the decreasing contribution of the
diffusion
broadening. At small detuning $|\Omega |\ll kv_T$ we have the explicit
expression
$$
D\simeq {\frac{\Gamma kv_T}{\sqrt{\pi }}}\left( 1+\left(
{\frac \Omega {kv_T}}\right) ^2\right) +
{\frac{\nu (kv_T)^2}{2\Gamma _j}}\left( {\frac{3\pi }2}-3
-\left( {\frac \Omega {kv_T}}\right) ^2\left( 15-{\frac{9\pi
}2}\right)
\right) .
$$

Summarizing the results, we conclude that the Bennett hole narrows
down as
detuning increases. The physical reason of the effect is that the
width of
the hole is not affected by the homogeneous width $\Delta v_H$,
but depends on the change in velocity due to diffusion $\Delta v_D =
v_T \sqrt{\nu\tau_j}$, while $\Delta v_D\gg \Delta v_H$. The excited
state lifetime $\tau_j=\Gamma _j^{-1}$ determines the stationary
width of the hole.  While the detuning $\Omega $ of the field remains
inside the Doppler contour, the field interacts only with a group of
slow ions $|v_z|\ll v_T$.  When we turn the field off the resonance,
the light interacts with faster excited ions. Collision frequency
of those ions with ions in the ground state $\nu$ is smaller,
therefore the width decreases. It would be more proper to call this
effect decrease of the broadening instead of the narrowing.

To observe the narrowing experimentally, the
probe-field spectroscopy can be used. One should measure the
near-resonant absorption or gain of a weak probe wave as a function
of its detuning on the same or adjacent transition. Another
possibility is recording the spectrum of spontaneous emission in the
presence of a strong resonant continuous field.  Let us estimate two
alternative effects independent of the detuning, namely, Stark
broadening and change of velocity due to the interaction with
electrons. The Stark effect is quadratic in the ArII spectrum. Its
value \cite[p.139]{Babin} is about 100~MHz, which is small compared
with both the homogeneous width $\Gamma $ and the diffusion one
$kv_T\sqrt{\nu /\Gamma _j}$.  The velocity change owing to scattering
on electrons is as small as the ratio of the ion and electron thermal
velocities $v_{Ti}/v_{Te}\sim 10^{-2}$ \cite[p.155]{Babin}. The
ion-ion collision frequency decreases no more than by a factor of
2--3 at the detuning about $kv_T$; therefore, neither alternative
broadening mechanism can compensate the effect in question. The
measured diffusion broadening in low-temperature argon plasma at zero
detuning is by a factor of 3--4 (see \cite[p.199]{Babin}). At the
detuning $\Omega \sim kv_T$ the narrowing factor expected is more
than 10\%, so this effect seems observable, see Fig. 1.

We thank S.A.Babin and S.G.Rautian for stimulating discussions,
E.G.Shapiro for excellent assistance in computation, and A.V.
Shafarenko for valuable comments. The research described in this
publication was made possible in part by Grant No. RCN000 from the
International Science Foundation.

\section*{List of captions}
Fig.1 (a) Distribution $\rho_j^{(1)}(v_z)$ of $j$-level population in
ion velocity $v_z$ within the first order of the perturbation theory
in the field intensity. Different curves correspond to distinct
values of detuning $\Omega$ (from left to right $\Omega/{kv_T} =
0; 0.5; 1; 1.5; 2; 2.5$). Parameters are assumed to be $\Gamma =
10^{-2} kv_T$, $\Gamma_j = 10^{-3} kv_T$, $\nu = 10^{-5} kv_T$.

(b) Half-width on half a maximum of contour $\rho_j^{(1)}(v_z)$ as a
function of detuning $\Omega/{kv_T}$ normalized to unit magnitude.


\begin{references}
\bibitem{Babin}  S.A.~Babin, D.A.~Shapiro, Phys.~Rep. {\bf 241}, 119
(1994).

\bibitem{Berman}  P.R.~Berman, Adv. At. Mol. Phys. {\bf 13}, 57
(1977).

\bibitem{Shalagin} S.G.~Rautian, A.M.~Shalagin, Kinetic Problems of
Non-Linear Spectroscopy (Elsevier, Amsterdam, 1991).

\bibitem{Bennett62}  W.R.~Bennett, Jr., Phys.~Rev. {\bf 126}, 580
(1962);
Appl. Opt., Supplement No 1, 24 (1962).

\bibitem{Smirnov}  G.I.~Smirnov, D.A.~Shapiro, Zh.~Eksp. Teor.~Fiz.
{\bf 76}, 2084 (1979) [Sov. Phys. JETP {\bf 49}, 1054 (1979)].

\bibitem{Chernykh}  E.V.~Podivilov, A.I.~Chernykh, D.A.~Shapiro,
Zh.~Eksp.  Teor.~Fiz. {\bf 105}, 1214 (1994) [Sov. Phys. JETP {\bf
78}, 653(1994)].

\bibitem{Stenholm}  S.~Stenholm, Phys. Rep. {\bf 43}, 151 (1978).

\bibitem{Bowles}  J.~Bowles, R.~McWilliams, N.~Rynn, Phys.~Rev.~Lett.
{\bf 68}, 1144 (1992).

\bibitem{Rautian}  S.G.~Rautian, Zh. Eksp. Teor. Fiz. {\bf 51}, 1176
(1966) [Sov. Phys. JETP {\bf 24}, 788 (1967)].

\bibitem{Gelmedova}  L.A.~Gel'medova, D.A.~Shapiro, J.~Mod.~Opt.
{\bf 38}, 573 (1991).
\end{references}
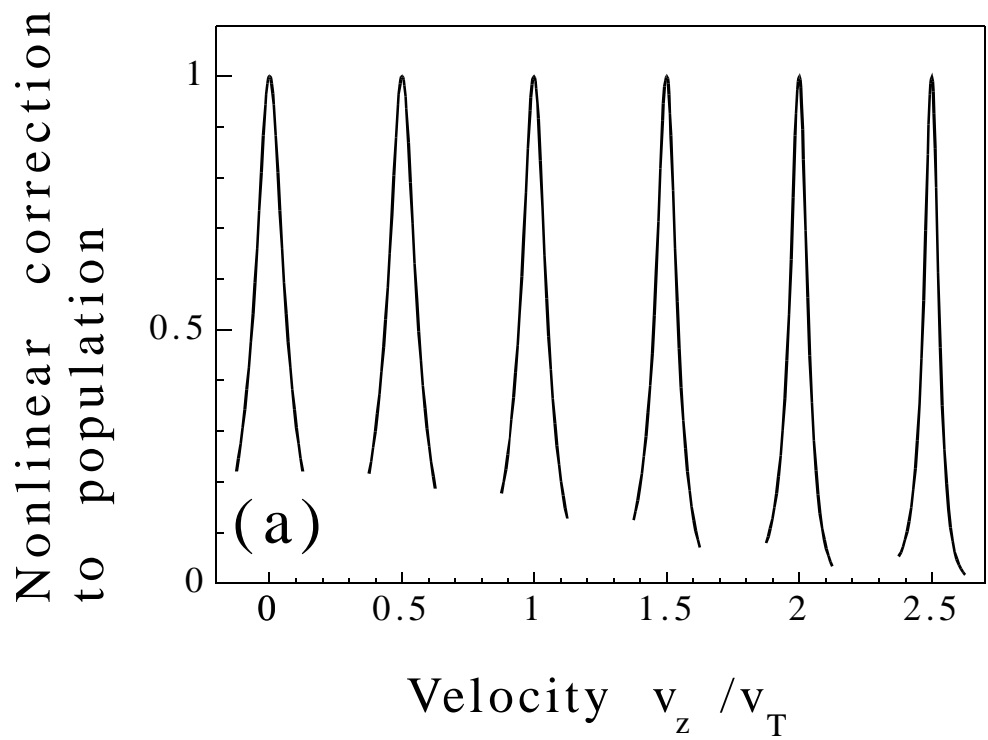
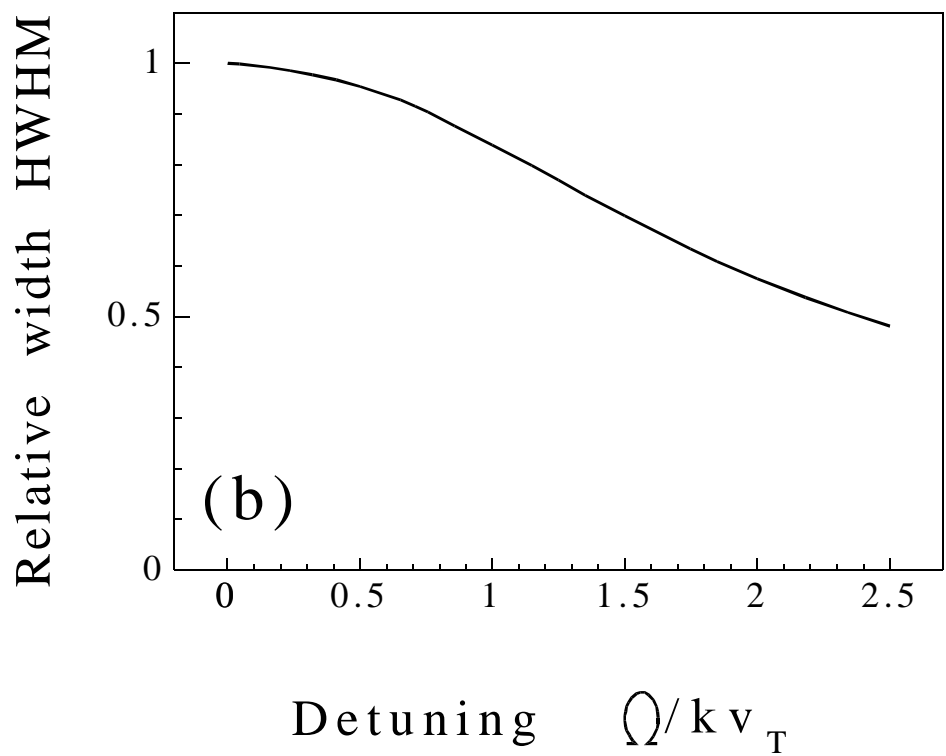
\end{document}